# Network analysis and Eurozone trade imbalances

**Giovanni Carnazza** (*corresponding author*) [(*)]

**Pierluigi Vellucci** [(*)]

## Abstract

European Monetary Union continues to be characterised by significant macroeconomic imbalances. Germany has shown increasing current account surpluses at the expense of the other member states (especially the European periphery). Since the creation of a single currency has implied the impossibility of implementing competitive devaluations, trade imbalances within a monetary union can be considered unfair behaviour. We have modelled Eurozone trade flows in goods through a weighted network from 1995 to 2019. To the best of our knowledge, this is the first work that applies this methodology to this kind of data. Network analysis has allowed us to estimate a series of important centrality measures. A polarisation phenomenon emerges in relation to the growth of German dominance. The common currency has then not been capable to remove trade asymmetry, increasing the distance between surplus and deficit countries. This situation should be addressed with expansionary policies on the demand side at national and supranational level.

[(*)] Università degli Studi Roma Tre, Department of Economics – giovanni.carnazza@uniroma3.it
[(*)] Università degli Studi Roma Tre, Department of Economics – pierluigi.vellucci@uniroma3.it

# 1. Introduction

Macroeconomic imbalances within the European Monetary Union (EMU) remain significant even though twenty years have passed since the beginning of the common currency area (1999). In this context, Germany – which has been considered for long the 1990s 'sick-man of Europe' – has rapidly become a dominant economic power (Gräbner et al., 2020). This is particularly true if we look at the progressive large accumulation of its current account surpluses. This characterization seems to contradict who had interpreted the common currency as a way to help the convergence among member states (Gill and Raiser, 2012): countries that have joined the EMU have progressively shown diverging trends in competitiveness which have contributed to the growth of significant current account imbalances, with northern European countries accumulating surpluses, and southern countries experiencing current account deficits.[1]

A large share of trade imbalances derives from trade with countries outside the euro area; nevertheless – and based on the finding that trade imbalances are in any case significant within the EMU – we mainly focus on intra-Eurozone trade flows to deepen how the sudden, but planned, repricing of the nominal exchange rate in favour of a single currency has eventually modified the relative positions among member states. At this regard, the magnitude of internal trade imbalances has been historical estimated and confirmed, applying Frisch's (1947) matrix to EMU market. In this context, trade in goods within the Eurozone has been modelled through a weighted directed network, taking into consideration a period covering 1995 to 2019. On the one side, the year '1995' will represent our baseline scenario and will be considered as a proxy of the pre-euro trade situation; on the other side, the year '2019' will be the year of comparison in relation to the effects of twenty years of the single currency. In any case, our analysis covers the entire period and remains robust regardless of the period considered.[2] In this way, we are free from any worries about the health and the subsequent economic crisis happened in the beginning of 2020. Network analysis allows us to estimate a series of important centrality measures to identify the most influential countries in the Eurozone network, assigning numbers and rankings to countries corresponding to their relative Eurozone network position. In this context, we will refer to the network of member states within the EMU as *EMU network*.

A social network can be defined as a set of actors, where each one of them is characterized by a connection to some or all other actors. The concept is broad enough that it can be applied to several phenomena (Worrell et al., 2013): an actor can be a single person, a company, or a country while the link can be a friendship relation between two people, a business relationship among companies or a trade relation between countries. Recently, this kind of analysis has received great attention with the aim of characterizing links' patterns, interactions, and implications of the relationships present in social networks. From an empirical point of view, network analysis has become very popular in the finance field, mostly after the Lehman Brothers' bankruptcy of 2008 (Cimini, 2015; Deev and Lyócsa, 2020). In particular, it has been mainly used to investigate interconnections among financial institutions and to forecast possible financial crises (Brunetti et al., 2019), channels of contagion, shock propagation and systemic risk (Lyócsa et al., 2017; Silva et al., 2018). The complex network theory has also become popular in the field of international trade (An et al., 2014; Fan et al., 2014; Zhang et al. 2014; Zhong et al. 2014; Vidmer et al. 2015; Ge et al., 2016; Nuss et al., 2016; Tokito et al., 2016). In this context, nodes represent the countries and links tend to be represented by the cash flow in the trade transactions. For example, De Andrade and Rêgo (2018) apply this methodology to the international trade network to observe which countries are the most connected in the world economy and are central to the intermediation of the wealth flow. Anyway, to the best of our knowledge, this is the first paper which applies this methodology to this kind of data, trying to quantify the relative importance and influence of member states within the EMU. In other words, our paper tries to address a relevant gap in the existing literature by applying network theoretical analysis to the EMU-trade network. Unlike other works in this field which tend to focus their attention on the methodological aspects of the analysis, we have applied two different procedures of normalizations to show the results in a more meaningful way from an economic point of view. This has allowed us to interpret our outcomes comparing 2019 with 1995, deepening the relative variations and the possible reasons behind them.

---

[1] Northern countries are also called 'core countries' while Southern countries are labelled as 'periphery countries'. Austria, Belgium, Finland, France, Germany, and the Netherlands are generally considered core countries while the periphery group is commonly composed of Greece, Ireland, Italy, Portugal, and Spain (De Santis and Cesaroni, 2016). However, other authors tend to differentiate countries taking into consideration specific indicators; for example, on the basis of GDP per capita, Piton (2021) adds Italy as a core country together with Luxembourg.

[2] The Appendix reports all the centrality measures for each country and for each year (see Tables from A.1 to A.9). As we will better explain later, different standardisations have been applied to show various aspects of the EMU network.

Our paper essentially answers three interrelated research questions. First, how is the EMU-trade network structured and how is it changed over time? Secondly, has the common currency reduced trade asymmetry within the EMU, fostering a better internal balance between deficit and surplus countries? Finally, which countries occupy the most influential positions and have these positions progressively modified? In the latter regard, the general finding that Germany is central in the EMU-trade network could appear trivial, being the largest European economy. However, at least two aspects should be considered: on the one side, this outcome should be also contextualised in dynamic terms with Germany that has significantly increased its relative weight in the EMU network[3]; on the other side, this positive trend has been associated with an increase in trade asymmetry among member states, which highlights an important stylized fact of the EMU trade flows.

---

[3] The deepening of the reasons behind the growing accumulation of German trade surpluses are beyond the scope of the present work. In any case, some explanations that have been introduced in economic literature are analysed in Section 2, which sets the main issue of the EMU trade imbalances.

The rest of the paper is structured as follows. Section 2 introduces the issue of trade imbalances within the EMU, highlighting some possible explanations progressively advanced in the economic literature. Section 3 introduces the methodology and how data have been processed to fit the network analysis; the main characteristics of this kind of analysis are then described as well as the main centrality measures used to investigate the EMU network and its changes over time. Section 4 concludes and provides some policy implications driven by the results.

## 2. Setting the issue: the EMU trade imbalances

One of the main causes of the euro crisis tends to be attributed to a balance-of-payments problem: persistent German current account surpluses have progressively translated into persistent deficits of countries in the European periphery (Simonazzi et al., 2013). At this regard, Figure 1 shows the significant increase of the German current account surplus during the 2000s, where the crisis of 2009 brought only a slight decrease in the surplus without influencing the overall positive trend. These imbalances have been generally justified by two different explanations.

The first explanation interprets the German surplus as an expression of a 'virtuous' saving behaviour that should be also extended to the periphery. In this sense, the surplus is seen as a reflection of an improvement of price competitiveness of German industry to the detriment of peripheral countries. In particular, the loss of competitiveness of the latter countries is often associated to the fall in borrowing costs following the creation of the currency area, which led to a remarkable increase of domestic demand, inflation, and then relative prices. These dynamics related to the rising external imbalances of Eurozone debtor countries can be traced back to two traditional and complementary justifications, which rely primarily on intra-Eurozone factors: firstly, financial integration and expectations of convergence of interest rates; secondly, over-optimism and wage/price rigidities. This represents the so-called 'intra-Eurozone competitiveness' problem: reduced borrowing costs and optimistic expectations of convergence – due to the elimination of currency risk – have led to credit booms with increases in domestic prices and demand and with labour costs inconsistent with the underlying productivity; as a result, in this kind of countries, labour costs have grown faster than in other countries, such as Germany (Chen et al., 2013).[4] In this context, it is important to keep in mind that trade within a currency union is significantly more reactive to changes in relative prices than extra-trade. This feature derives from the substantial difficulty to adjust relative prices to restore competitiveness, since the single currency does not allow for competitive devaluations (Bayoumi et al., 2011).

---

[4] Jaumotte and Sodsriwiboon (2010) conducted a study on the drivers of current account deficits, finding that medium-term fundamentals – such as demographic trends, the level of development relative to trading partners and relative fiscal positions – do not play an important role. On the contrary, the adoption of the euro has been found as a significant determinant.

The second explanation behind the increase of the German current account surplus during the 2000s is based on two common mistakes: on the one side, the idea of a universal replication of an export-led growth model; on the other side, the underestimation that high net exports are linked to a stagnant domestic demand mainly due to wage compression (Whyte, 2010). These two beliefs tend to respectively misinterpret two economic arguments: first of all, chronic weak demand and large external surpluses are sustainable only if other countries behave in the opposite way; secondly, competitiveness should not be confused with productivity, being current account surpluses a reflection of Germany's domestic general weakness (Simonazzi et al., 2013).

**Figure 1 – Germany: current account balance - % of GDP**

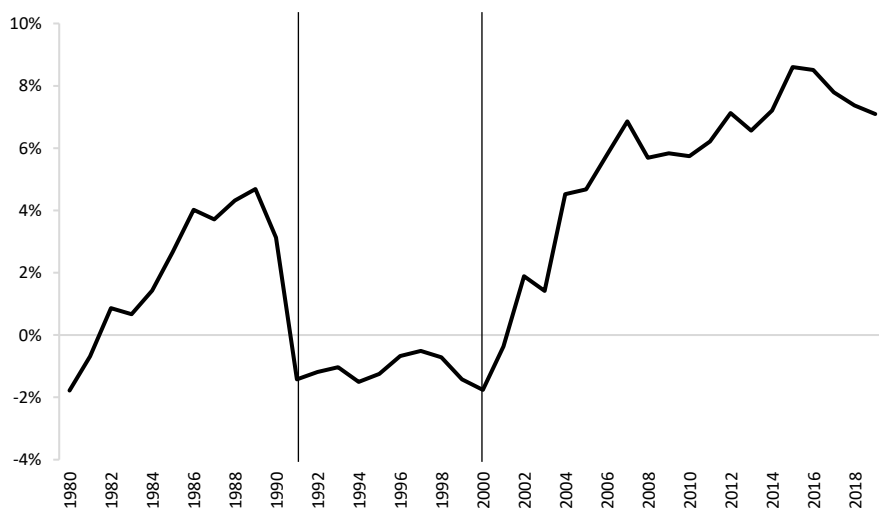

*Source:* elaborations on *IMF-WEO* data

From a quantitative point of view, since the introduction of the euro, pay restraint in Germany has been quite exceptional: pay settlements systematically undershoot the rate of productivity growth, determining a consistent fall in real unit labour costs. Consequently, Germany's real effective exchange rate (REER) has been characterized by a marked decline and this negative trend can be considered a central factor behind the consequent rising share of world exports.[5] At this regard, taking into consideration 2000 and 2019, it is possible to point out that Germany is the only country – together with Cyprus – where the REER has recorded practically the same value after twenty years of adoption of the single currency; all the other countries have been characterized from a not negligible increase (Figure 2).[6] Movements of the REER can be generally decomposed into two components: on the one side, movements in domestic prices (or unit labour costs) relative to those of trading partners; on the other side, movements of the nominal exchange rate. From this point of view, in Germany the nominal appreciation has been generally offset by a decline in unit labour costs relative to trading partners. In the first decade of the EMU, the initial euro appreciation has reflected a significant inflow of capital directed not only to the Eurozone periphery but above all to core Eurozone countries, such as France and Germany: in this sense, at least part of the ensuing loss in competitiveness can be considered an external shock for countries in the Eurozone periphery (Chen et al., 2013).

---

[5] The different abilities of countries to export their goods depend not only on costs but also on technological competitiveness. At this regard, the accumulation of technological capabilities can be considered another important driver of export competitiveness. This kind of driver tends to be path dependent, which could represent a significant feature in explaining Germany's persistent position as an export leader in the EMU and the subsequent difficulties for other countries to copy this development model (Dosi et al., 2015; Aistleitner et al., 2021).

**Figure 2 – Real effective exchange rates (REER) within the Eurozone (2010 = 100)**

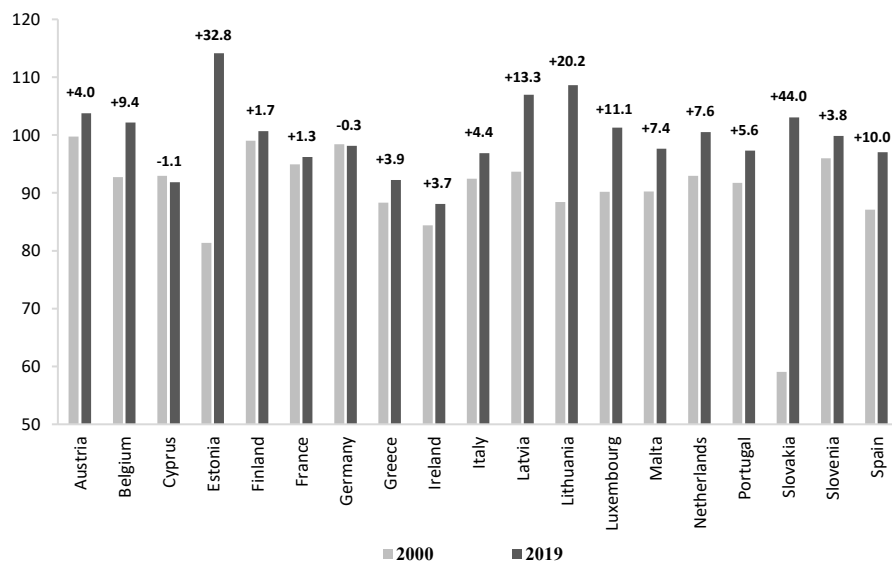

*Note:* the consumer price index has been used as deflator. The REER shows the national exchange rate against a basket of currencies, adjusted for the effects of inflation. In this case, it has been calculated in relation to 42 trading industrial countries. Above the column of each country is reported the change in percentage points between 2000 and 2019.
*Source:* elaborations on *Eurostat* data

The decline in labour costs relative to trading partners which has characterized Germany since the beginning of the EMU can be clarified taking into consideration two aspects.[7] First of all, when the German Democratic Republic became part in 1990 of the Federal Republic of Germany to form the reunited nation of Germany, real wages significantly increased due to the jump of wages in east Germany and a mainly financed construction boom (Coricelli and Wörgötter, 2012). This dynamic has damaged competitiveness of the German export sector, and it is behind the current account deficits which have been recorded during the 1990s (Figure 1). The subsequent fall – and then the substantial immobility – in real wages derives from a series of reforms that have increased flexibility (part-time and temporary jobs) in the labour market (Figure 2). These reforms determined wage moderation and a consequent growth in profits and competitiveness of German firms.[8]

---

[6] Table A.10 in the Appendix reports different measures of the REER, which highlight the significant reduction occurred in Germany between 1995 and 2000 and the overall decrease between 2000 and 2019.
[7] The aforementioned decline in labour costs needs also to be contextualised within a general increasing path of unit labour costs in periphery countries since the mid-1990s. At this regard, Piton (2021) analyses this divergence between core and peripheral countries by highlighting two different historical periods: at first, this dynamic was interpreted as reflecting catching-up processes in the periphery (Blanchard and Giavazzi, 2002); subsequently – mainly after the financial crisis of 2008 and the sovereign debt crisis – ULC divergence has been considered a fundamental amplifying factor in the crisis (Shambaugh, 2012). For more information about the origins of ULC divergence, see among others Piton (2021).

The previous two explanations behind German trade surplus are relevant because they try to justify the emergence and persistence of these trade imbalances within the Eurozone. From an empirical point of view, it is also important to account for other aspects that have characterized the remarkable growth of Germany's trade surplus. First, Germany has been able to integrate its production chain with the progressive enlargement of the single European market towards eastern European countries. This eastward expansion was the prerogative of countries, such as Germany, which took advantage of cultural ties and closer borders and has gone together with an impoverishment of the productive matrix of those southern regions less connected with Germany and with the general redirection of trade flows (Simonazzi et al., 2013). In this context of elimination of barriers to trade and investments, Germany has outsourced only parts of its manufacturing activity (low- and high- skill jobs) while activities characterized by medium skills have remained in the country (Coricelli and Wörgötter, 2012). This kind of delocalization has contributed to the creation of jobs in the home country, determining at the same time a significant reduction of unit labour costs (Marin, 2010a, 2010b). Secondly, the growth of the Chinese – and more generally emerging Asian – economy has increased the demand for those goods, such as machinery and equipment, which were mainly exported by Germany. Finally, a similar dynamic can be found in the rise of oil prices: fast income growth in commodity-exporting countries may have benefited countries such as Germany exporting goods in high demand by oil producers (Chen et al., 2013).

In this context, great importance has been attributed to the 'upgrading' of German exports while less attention has been paid to the related 'downgrading' of imports. From this point of view, it is important to take into consideration the linkage between imports and income distribution. As previously seen, in Germany real wages have remained substantially constant in the last twenty years (Figure 2). In particular, phenomena such as the spread of low-paid jobs and unemployment contributed to trade surplus in two different ways: on the one side, they directly weakened domestic consumption and therefore the imports of goods; on the other side, they redirected consumption from European goods towards lower-quality products imported from emerging countries – in other words,

---

[8] The same labour market reforms may have encouraged job insecurity and uncertainty among works, determining a consequent increase in precautionary savings and a fall in consumption (Bertola and Lo Prete, 2011).

low- price imports (especially from China) displaced middle- and higher- price products from the Eurozone countries (Simonazzi et al., 2013).

Shifting our focus on the Eurozone as a whole, it seems clear that the single market has not been able to eliminate – or at least to reduce – trade imbalances. On the contrary, the elimination of foreign exchange risk premia has helped to develop large current account imbalances within the Eurozone (Waysand et al., 2010). At this regard, Jaumotte and Sodsriwiboon (2010) highlight an interesting aspect related to the introduction of the euro: current account deficits have been fostered by allowing countries to invest more than what could be financed from national saving (economic integration improved access to the international pool of saving). Moreover, the lack of an adjustable nominal exchange rate has resulted in more persistence trade imbalances and this conclusion is particularly true for those countries characterized by high levels of employment protection.[9] Market institutions seem to affect significantly external balances, and this is the reason why Berger and Nitsch (2010) consider the latter characterization as one of the main causes of trade imbalances within the Eurozone.

The magnitude of trade imbalances can be estimated by looking at the relative skewness of the trade matrix as initially defined by Frisch (1947) and measured as the ratio of the sum of the surpluses (absolute skewness) – which, within a closed sample, is equal to the sum of the deficits – to total trade (total exports/imports). In order to better explain the important features of the Frisch's trade matrix and how the relative skewness is estimated, we first clarify how it is built (Table 1). Taking into consideration EMU member countries, each row represents an exporting country and the related bilateral export while each column denotes an importing country and the related bilateral import (countries are placed in the same order). If the sum of the elements in a row is larger than the sum of the elements in the corresponding column, the country in question is defined as a surplus country (i.e., Germany in both years); in the opposite case, the country is characterized by a trade deficit (i.e., Austria in both years). As noted above, in a closed trade matrix the sum of the exports must be equal to the sum of the imports and this amount represents the overall trade (873 billion dollars in 1995 and 2,007 billion dollars in 2019). The same feature applies to the sum of the surpluses and of the deficits. This quantity defines the absolute skewness (63 billion dollars in 1995 and 186 billion dollars in 2019); in our case, the absolute skewness can be interpreted as an expression for the amount of liquid transfers or international lending that is needed. The ratio between the absolute skewness and the overall trade defines the concept of relative skewness within a closed sample (7.3% in 1995 and 9.3% in 2019). From a theoretical point of view, when facing exclusively non-negative numbers – as in the case of a trade matrix – the relative skewness must be a number between 0 and 100 per cent; it is zero

---

[9] The question whether exchange rate variability affects the speed of current account adjustment can be traced back to Friedman (1953), who firstly claimed that flexible exchange rates allow for prompt and continuous change of relative prices and thereby facilitate rapid external adjustment.

## Table 1 – Frisch's trade matrix within the Eurozone (billion dollars)

### (a) 1995

| | | AT | BE | CY | EE | FI | FR | DE | EL | IE | IT | LV | LT | LU | MT | NL | PT | SK | SL | ES | Total | Surplus |
|---|---|---|---|---|---|---|---|---|---|---|---|---|---|---|---|---|---|---|---|---|---|---|
| Exporting countries | AT | - | 0.9 | 0.0 | 0.0 | 0.4 | 2.4 | 20.2 | 0.3 | 0.1 | 4.8 | 0.0 | 0.0 | 0.1 | 0.0 | 1.6 | 0.2 | 0.6 | 1.0 | 1.1 | 34 | |
| | BE | 1.6 | - | 0.1 | 0.1 | 0.9 | 27.9 | 32.0 | 0.9 | 0.6 | 8.9 | 0.0 | 0.1 | 4.1 | 0.1 | 20.7 | 1.1 | 0.2 | 0.2 | 4.5 | 104 | 14 |
| | CY | 0.0 | 0.0 | - | 0.0 | 0.0 | 0.1 | 0.1 | 0.1 | 0.0 | 0.0 | 0.0 | 0.0 | 0.0 | 0.0 | 0.0 | 0.0 | 0.0 | 0.0 | 0.0 | 0 | |
| | EE | 0.0 | 0.0 | 0.0 | - | 0.4 | 0.0 | 0.2 | 0.0 | 0.0 | 0.0 | 0.1 | 0.1 | 0.0 | 0.0 | 0.1 | 0.0 | 0.0 | 0.0 | 0.0 | 1 | |
| | FI | 0.4 | 1.2 | 0.0 | 0.9 | - | 1.9 | 5.4 | 0.2 | 0.2 | 1.2 | 0.2 | 0.1 | 0.0 | 0.0 | 1.7 | 0.2 | 0.1 | 0.0 | 1.1 | 15 | 4 |
| | FR | 3.3 | 22.3 | 0.2 | 0.0 | 1.2 | - | 48.7 | 2.2 | 1.5 | 27.6 | 0.0 | 0.1 | 1.5 | 0.3 | 12.6 | 4.0 | 0.3 | 0.8 | 20.5 | 147 | |
| | DE | 27.4 | 28.2 | 0.5 | 0.3 | 4.7 | 58.4 | - | 3.8 | 2.5 | 38.5 | 0.4 | 0.5 | 2.9 | 0.3 | 38.0 | 4.7 | 2.1 | 2.2 | 17.4 | 233 | 7 |
| | EL | 0.2 | 0.2 | 0.3 | 0.0 | 0.1 | 0.7 | 2.4 | - | 0.0 | 1.5 | 0.0 | 0.0 | 0.0 | 0.1 | 0.3 | 0.1 | 0.0 | 0.0 | 0.4 | 6 | |
| | IE | 0.3 | 1.8 | 0.0 | 0.0 | 0.3 | 3.7 | 5.8 | 0.3 | - | 1.9 | 0.0 | 0.0 | 0.0 | 0.0 | 2.4 | 0.2 | 0.0 | 0.0 | 1.2 | 18 | 11 |
| | IT | 5.5 | 6.3 | 0.4 | 0.1 | 1.1 | 29.9 | 42.3 | 4.4 | 0.9 | - | 0.1 | 0.1 | 0.3 | 1.2 | 6.8 | 3.2 | 0.5 | 1.9 | 11.2 | 116 | 9 |
| | LV | 0.0 | 0.0 | 0.0 | 0.1 | 0.1 | 0.0 | 0.3 | 0.0 | 0.0 | 0.0 | - | 0.1 | 0.0 | 0.0 | 0.1 | 0.0 | 0.0 | 0.0 | 0.0 | 1 | |
| | LT | 0.0 | 0.0 | 0.0 | 0.1 | 0.0 | 0.1 | 0.4 | 0.0 | 0.0 | 0.1 | 0.2 | - | 0.0 | 0.0 | 0.2 | 0.0 | 0.0 | 0.0 | 0.0 | 1 | |
| | LU | 0.1 | 1.1 | 0.0 | 0.0 | 0.0 | 1.1 | 1.8 | 0.1 | 0.0 | 0.6 | 0.0 | 0.0 | - | 0.0 | 0.4 | 0.1 | 0.0 | 0.0 | 0.2 | 6 | |
| | MT | 0.0 | 0.1 | 0.0 | 0.0 | 0.0 | 0.3 | 0.3 | 0.0 | 0.0 | 0.7 | 0.0 | 0.0 | 0.0 | - | 0.0 | 0.1 | 0.0 | 0.0 | 0.0 | 2 | |
| | NL | 2.5 | 24.1 | 0.1 | 0.1 | 1.3 | 16.8 | 43.1 | 1.8 | 1.1 | 10.5 | 0.1 | 0.1 | 0.5 | 0.1 | - | 1.5 | 0.2 | 0.2 | 4.9 | 109 | 19 |
| | PT | 0.3 | 0.7 | 0.0 | 0.0 | 0.2 | 3.3 | 5.0 | 0.1 | 0.1 | 0.9 | 0.0 | 0.0 | 0.0 | 0.0 | 1.2 | - | 0.0 | 0.0 | 3.5 | 15 | |
| | SK | 0.5 | 0.1 | 0.0 | 0.0 | 0.0 | 0.2 | 1.7 | 0.0 | 0.0 | 0.5 | 0.0 | 0.0 | 0.0 | 0.0 | 0.2 | 0.0 | - | 0.1 | 0.1 | 3 | |
| | SL | 0.6 | 0.1 | 0.0 | 0.0 | 0.0 | 0.7 | 2.6 | 0.0 | 0.0 | 1.3 | 0.0 | 0.0 | 0.0 | 0.0 | 0.1 | 0.0 | 0.1 | - | 0.1 | 6 | |
| | ES | 0.8 | 2.6 | 0.1 | 0.0 | 0.4 | 18.4 | 13.6 | 0.9 | 0.3 | 8.1 | 0.0 | 0.0 | 0.1 | 0.1 | 3.4 | 7.3 | 0.1 | 0.2 | - | 57 | |
| | Total | 44 | 90 | 2 | 2 | 11 | 166 | 226 | 15 | 8 | 107 | 1 | 1 | 10 | 2 | 90 | 23 | 4 | 7 | 66 | 873 | 63 |
| | Deficit | -10 | | -1 | -1 | | -19 | | -9 | | | | | -4 | -1 | | -7 | -1 | -1 | -10 | -63 | **7.3%** |

### (b) 2019

| | | AT | BE | CY | EE | FI | FR | DE | EL | IE | IT | LV | LT | LU | MT | NL | PT | SK | SL | ES | Total | Surplus |
|---|---|---|---|---|---|---|---|---|---|---|---|---|---|---|---|---|---|---|---|---|---|---|
| Exporting countries | AT | - | 3.2 | 0.1 | 0.2 | 0.7 | 7.4 | 48.4 | 0.6 | 0.4 | 11.0 | 0.2 | 0.4 | 0.2 | 0.1 | 3.3 | 0.5 | 3.5 | 3.6 | 2.7 | 86 | |
| | BE | 3.9 | - | 0.3 | 0.4 | 2.1 | 55.6 | 68.9 | 1.9 | 2.7 | 22.1 | 0.4 | 0.9 | 6.9 | 0.2 | 53.1 | 2.8 | 1.2 | 0.7 | 10.8 | 235 | 21 |
| | CY | 0.0 | 0.0 | - | 0.0 | 0.0 | 0.1 | 0.1 | 0.4 | 0.0 | 0.2 | 0.0 | 0.0 | 0.0 | 0.1 | 0.1 | 0.0 | 0.0 | 0.0 | 0.0 | 1 | |
| | EE | 0.1 | 0.3 | 0.0 | - | 2.3 | 0.4 | 1.0 | 0.0 | 0.0 | 0.2 | 1.4 | 1.1 | 0.0 | 0.0 | 0.6 | 0.0 | 0.0 | 0.0 | 0.2 | 8 | |
| | FI | 0.5 | 2.6 | 0.1 | 1.8 | - | 2.6 | 10.4 | 0.2 | 0.2 | 2.7 | 0.7 | 0.7 | 0.0 | 0.0 | 4.5 | 0.2 | 0.2 | 0.1 | 1.4 | 29 | |
| | FR | 4.5 | 39.0 | 0.3 | 0.5 | 2.2 | - | 76.2 | 2.7 | 4.4 | 41.2 | 0.4 | 0.9 | 2.8 | 0.8 | 19.6 | 8.3 | 3.3 | 1.5 | 40.8 | 249 | |
| | DE | 67.5 | 51.3 | 0.7 | 1.9 | 11.7 | 114.6 | - | 6.6 | 8.8 | 75.1 | 1.8 | 3.8 | 6.0 | 0.9 | 87.7 | 11.7 | 15.2 | 5.9 | 47.8 | 519 | 45 |
| | EL | 0.3 | 0.5 | 2.0 | 0.0 | 0.1 | 1.5 | 2.5 | - | 0.2 | 3.7 | 0.0 | 0.1 | 0.0 | 0.3 | 0.8 | 0.2 | 0.2 | 0.6 | 1.1 | 14 | |
| | IE | 0.6 | 19.0 | 0.1 | 0.1 | 0.5 | 8.1 | 18.5 | 1.0 | - | 5.0 | 0.1 | 0.1 | 0.1 | 0.1 | 9.9 | 0.6 | 0.2 | 0.2 | 3.1 | 67 | 41 |
| | IT | 11.7 | 15.9 | 1.2 | 0.5 | 1.9 | 54.9 | 64.5 | 5.2 | 2.0 | - | 0.6 | 1.3 | 0.7 | 1.8 | 13.4 | 4.8 | 3.5 | 5.2 | 26.7 | 216 | |
| | LV | 0.1 | 0.2 | 0.0 | 1.5 | 0.3 | 0.3 | 1.1 | 0.0 | 0.1 | 0.2 | - | 2.4 | 0.0 | 0.0 | 0.4 | 0.0 | 0.1 | 0.0 | 0.2 | 7 | |
| | LT | 0.2 | 0.6 | 0.0 | 1.5 | 0.6 | 0.8 | 2.4 | 0.1 | 0.1 | 0.8 | 2.9 | - | 0.0 | 0.0 | 1.2 | 0.1 | 0.1 | 0.1 | 0.4 | 12 | |
| | LU | 0.4 | 1.8 | 0.0 | 0.0 | 0.1 | 2.1 | 3.7 | 0.0 | 0.1 | 0.8 | 0.0 | 0.1 | - | 0.0 | 1.0 | 0.1 | 0.1 | 0.1 | 0.5 | 11 | |
| | MT | 0.0 | 0.0 | 0.0 | 0.0 | 0.0 | 0.4 | 0.6 | 0.0 | 0.0 | 0.4 | 0.0 | 0.0 | 0.0 | - | 0.1 | 0.0 | 0.0 | 0.0 | 0.2 | 2 | |
| | NL | 6.1 | 65.7 | 0.4 | 0.8 | 4.3 | 37.0 | 106.4 | 2.9 | 4.7 | 23.9 | 0.7 | 1.7 | 1.4 | 0.3 | - | 4.3 | 1.7 | 0.9 | 16.5 | 280 | 68 |
| | PT | 0.7 | 1.6 | 0.1 | 0.0 | 0.4 | 8.5 | 8.1 | 0.2 | 0.5 | 2.8 | 0.1 | 0.1 | 0.1 | 0.1 | 2.7 | - | 0.5 | 0.2 | 15.4 | 42 | |
| | SK | 4.8 | 1.3 | 0.1 | 0.1 | 0.3 | 6.2 | 19.7 | 0.5 | 0.2 | 3.9 | 0.2 | 0.2 | 0.1 | 0.0 | 1.9 | 0.3 | - | 0.7 | 2.5 | 43 | 11 |
| | SL | 2.7 | 0.5 | 0.0 | 0.1 | 0.1 | 2.1 | 7.1 | 0.2 | 0.1 | 4.2 | 0.0 | 0.1 | 0.0 | 0.0 | 0.7 | 0.2 | 0.6 | - | 0.7 | 19 | |
| | ES | 2.7 | 9.7 | 0.4 | 0.3 | 1.1 | 47.4 | 34.9 | 2.5 | 1.9 | 25.5 | 0.3 | 0.6 | 0.4 | 0.4 | 10.9 | 25.8 | 1.4 | 0.7 | - | 167 | |
| | Total | 107 | 213 | 6 | 10 | 29 | 350 | 474 | 25 | 26 | 223 | 10 | 15 | 19 | 5 | 212 | 60 | 32 | 20 | 171 | 2,007 | 186 |
| | Deficit | -21 | | -5 | -2 | | -100 | | -11 | | -7 | -3 | -3 | -8 | -3 | | -18 | | -1 | -4 | -186 | **9.3%** |

*Source:* elaborations on *Observatory of Economic Complexity* (*OEC*) data

when and only when trade flows are completely balanced, that is when each country's exports equal its imports. Figure 3 illustrates the movements over time of the information seen in Table 1 – overall trade, absolute skewness, and relative skewness in relation to the EMU network from 1995 to 2019 – highlighting some interesting aspects.[10]

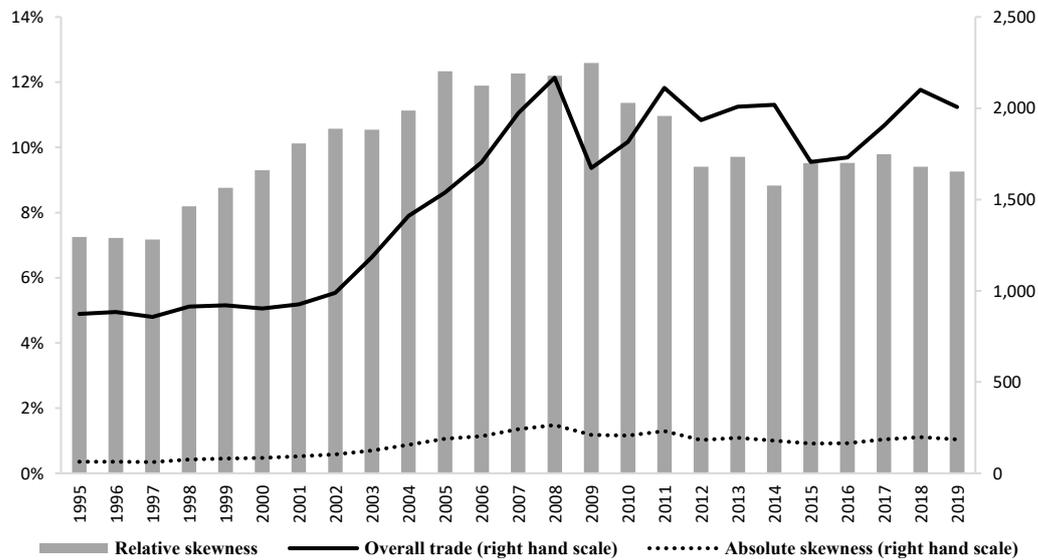

**Figure 3 – Absolute and relative skewness within the Eurozone (billion dollars)**

*Source:* elaborations on *Observatory of Economic Complexity* (*OEC*) data

First, the creation of the EMU seems to have been able to foster the overall exchange of goods: this amount has almost doubled over twenty years, signaling the ability of a single market to expand its internal trade. At this regard, the overall trade remains almost stable in the second half of the Nineties while the most significative increase occurs between 2002 and 2008, where the highest absolute values is recorded (2,167 billion dollars). Subsequently, the financial crisis of 2008 and the sovereign debt crisis led to a double contraction, before reaching pre-crisis levels again in 2019. Secondly, this remarkable growth did not match with a reduction of trade imbalances. On the contrary, the absolute skewness has grown at a higher rate until 2008, determining a progressive increase of the relative skewness (from 7.3% in 1995 to 12.2% in 2008). In this sense, it is possible to state that the single market enhanced – rather than discouraged – trade imbalances. Third, twenty years of monetary integration have not significantly changed the balance between surplus and deficit countries within the EMU: Belgium, Germany, Ireland, and the Netherlands have always displayed a surplus balance towards the other countries of the monetary union. Italy represents an interesting case being characterized by an overall trade surplus – apart from the period between the two crisis – mostly driven by exports with countries outside the EMU; intra-EMU net exports have displayed a negative trend since the beginning of the adoption of the common currency (Figure A.3 in the Appendix). In economic literature, such heterogeneous behavior has been also called 'Eurozone current account core-periphery dualism' (De Santis and Cesaroni, 2016) and our analysis confirms that it has been fostered by the intensification of European economic and financial integration process. As theoretically predicted by Frisch (1947), it has been probably illusory to believe that lower tariffs would have automatically led to a trading system both multilateral and able to eliminate – or reduce – trade imbalances (Simonazzi et al., 2013).

---

[10] In the Appendix, Figures A.4 and A.5 show the overall trade, the absolute skewness, and the relative skewness in relation to a subset of countries. Figure A.4 includes the 11 initial founders of the EMU while Figure A.5 adds Greece, which entered the monetary union two years later.

## 3. Network analysis and centrality measures

In this study, trade flows among member states of the EMU can be abstracted as a network of nodes and links: on the one side, the nodes represent the countries, and the links represent the relationships between each of the countries; on the other side, the direction of the links corresponds to the direction of the commercial flow of goods (i.e., imports and exports). EMU trade flow data come from the Observatory of Economic Complexity (OEC), which integrates and distributes data from a variety of official sources. In particular, we have taken into consideration all the 19 eurozone countries, tracing the complete trade network among individual countries. Data availability ranges from 1995 to 2019 and this represents our period of analysis. Unfortunately, from 1995 to 1998 data relative to Belgium and Luxembourg are presented in aggregate and this forced us to formulate some hypothesis to break down the two countries and the related trade flows.[11] This has become necessary in order to build a complete and consistent network with the remaining years.

*3.1 Methodology*

In mathematics, a graph is a structure made of vertices and edges; the vertices are also called nodes (or points) and are connected to each other by an edge (or link). Formally, it is possible to define an undirected graph as a pair $G = (V, E)$; $V$ is the set of vertices and $E$ is the set of edges, which are unordered pairs of elements of V. Mathematically, given the set of vertices $V$, we can define $E$ in the following way:

$$E = \{\{x, y\} | x, y \in V, x \neq y\} \qquad (1)$$

---

[11] From 1995 to 1998, the 'Belgium-Luxembourg' country has been decomposed in terms of trade flows in two different ways. First, we computed the overall bilateral trade with each country from 1999 to 2019, estimating the average share of Belgium and Luxembourg on total bilateral export. This average share has been then applied to the aggregate data, keeping it unchanged in overall terms. Secondly, trade flows between Belgium and Luxembourg from 1995 to 1998 have been calculated by applying a linear trend to the actual bilateral data from 1999 to 2019. The differences with other kind of trends are negligible.

The formal definition of a directed graph is similar, the only difference is that the set E contains ordered pairs of elements of V. Graphically, this translates into fixing a direction of travel of the connections and therefore of arrows from one node to another.[12] Mathematically, this means working with the Cartesian product $V^2 = V \times V$, i.e.

$$E = \{(x,y) \in V^2 | x \neq y\} \quad (3)$$

Network theory is a part of graph theory (Otte and Rousseau, 2002): a network can be defined as a graph where vertices and edges have attributes (i.e., names, money…). In a way, without these attributes, a network (i.e., a graph) becomes a hypothetical structure that does not exist in the real world. Edges can have weights – numbers that in some applications may be a measure of the length of a route or the capacity of a line. In this case, we will refer to these as weighted networks or graphs. Similarly, we will talk about undirected network and directed network. For this reason, we will use a terminology closer to the network theory by calling the vertices, nodes, and the edges, links.

In this paper, we model the trade within the Eurozone through a weighted directed network. A graph – and then a network – can be represented by a particular square matrix, the adjacency matrix. In place (i, j) of the matrix there is a non-zero number if and only if there is a link in the graph that goes from the node *i* to the node *j* (hence this number will be the weight of the link), otherwise there is a 0. Formally, we are talking about the $n \times n$ matrix *A* indexed by *V*, whose (i, j)-entry for a directed graph is defined as

$$A_{ij} = \begin{cases} w_{ij} & if\ (i,j) \in E \\ 0 & if\ (i,j) \notin E \end{cases} \quad (4)$$

where $w_{ij}$ is the weight of link (i, j). With each link $l = (i,j)$ of $E$ let there be associated a real number $w(l) = w_{ij}$, called its weight. We assume that each node of the directed network represents a European country and the directed link between two trading countries represents a flow of money between them; the outward arrow indicates the exports, the incoming arrow indicates the imports, the weight indicates the amount of exports or imports in euros. This happens according to a temporal (dynamical) mechanism, which modifies the numerical attributes of links between countries (i.e., the flow of money between them). More in details, if country *i* imports some good from country *j* during the year *t*, then a link from *i* to *j* is drawn in the t-th snapshot of the graph, corresponding to a non-zero entry in the corresponding adjacency matrix. In our problem, the adjacency matrix has 0s on the diagonal because countries cannot export to / import from themselves. From now on we suppose that all graphs and measures are fixed at time *t* and, since the purpose of this article is to show their trend over time, in the results section we will specify in which instant in time (i.e., year) they have been calculated.[13]

---

[12] For a visual representation of graphs, see Figure B.1 in Appendix B.

*3.2 The indicators for analyzing the EMU network*

Let us now introduce the centrality measures that we will use to identify the most influential countries in the EMU network. The centrality measures are indices that allow us to understand what characterizes an important node and then an important country. In such a description, they assign numbers and rankings to countries corresponding to their Eurozone network position. From a conceptual point of view, the simplest one is the weighted degree centrality, which is related to the number of links entering or exiting a certain node. If we look at the outgoing links, then we are talking about the weighted out-degree centrality, representing the export side of a country. If $n$ denotes the number of countries in our problem, the weighted out-degree centrality of country/node $i$ can be defined as follows (Wasserman and Faust K, 1994):

$$od_i = \sum_{j=1}^{n} w_{ij} = \sum_{j=1}^{n} A_{ij} \qquad (6)$$

for $i = 1, 2, \cdots, n$ (i.e., it can be seen as a sum by columns of the entries of *A*). The weighted out-degree centrality captures the outreach of a country to the community. A high weighted out-degree centrality value indicates that a country $i$ exports a lot, aiming to reach all other nations with a certain pervasiveness (all are practically connected with everyone, but the network is with weight and the weight in our example tells how pervasive the influence of $i$ is). The weighted out-degree centrality, then, captures the level of engagement a country $i$ initiates with members of the community. If we look at the incoming links, then we are talking about weighted in-degree centrality, the import side of a country. Formally, the weighted in-degree centrality of country/node *j* can be defined as follows (Wasserman and Faust K, 1994):

$$id_j = \sum_{i=1}^{n} w_{ij} = \sum_{i=1}^{n} A_{ij} \qquad (7)$$

for $j = 1, 2, \cdots, n$ (i.e., it can be seen as a sum by rows of the entries of *A*). The weighted in-degree centrality measures the number of links others have initiated with a particular country. Countries with high weighted in-degree centrality gain attention to their markets among the community of European states who participate in the exchange. Countries' weighted in-degree centrality, thus, captures the community's engagement with them. Those with high weighted in-degree centrality scores can be thought as market hubs since others have exported in them.

---

[13] In Appendix B, we introduce a toy network made of five countries to illustrate this concept (Figure B.2). The countries are *a*, *b*, *c*, *d*, *e*, *f*, *g* and *h*. We see that *e* exports to *b* for an amount of 6 euros and imports from *b* and *a*, respectively, for a total of 3 and 30 euros. Trade is typically non-symmetrical in the sense that a country can import more than it exports, so the associated adjacency matrix will be non-symmetric – see the adjacency matrix of the toy network showed in (B.1).

How far apart are two countries that trade directly (i.e., adjacent in the network) with each other? We need to define the distance between two countries, in terms of the flow of money between them. Then it is reasonable to assume the distance between two nodes/countries $i$ and $j$ to be the smaller, the greater the value exchanged from them. Hence, we can define the length $l$ of a weighted path of our network as the inverse of its weight $l_{ij} = 1/w_{ij}$ (De Andrade and Rêgo, 2018). Obviously, since we have a directed network with a non-symmetric adjacency matrix, we will generally have that $l_{ij} \neq l_{ji}$. This leads to another particularly important concept, the *shortest path*. The shortest path between two nodes $i$ and $j$ of a graph is the path that connects these nodes and that minimizes the sum of the lengths of its constituent links. We assume here that the length of a link $(i,j)$ is defined by means the distance between the adjacent nodes $i$ and $j$ (any two nodes connected by a link are said to be adjacent). Here we have employed Dijkstra's algorithm for detecting the shortest path between two nodes (Cormen et al, 2009; De Andrade and Rêgo, 2018).

Based on shortest paths, there is the betweenness centrality, a measure of centrality in a graph (Wasserman and Faust K, 1994). The betweenness centrality of a country/node $i$ is given by the expression:

$$b_i = \sum_{j \neq i \neq k} \frac{\sigma_{jk}(i)}{\sigma_{jk}} \qquad (7)$$

for $i = 1, 2, \cdots, n$, where $\sigma_{jk}$ is the total number of shortest paths from node $j$ to node $k$ and $\sigma_{jk}(i)$ is the number of those paths that pass through $i$ (not where $i$ is an end point). Therefore, the betweenness centrality is a measure that captures a different type of importance: the extent to which a certain country lies on the shortest paths between other countries. In other words, it helps identify countries which play a bridging role in a network.

Still based on shortest paths, there is the closeness centrality (Bavelas, 1950; Sabidussi, 1966), another measure of centrality in a graph, which in our paper we define as

$$c_i = \frac{n}{\sum_{j=1}^{n} \mu_{ij}} \qquad (8)$$

for each country $i = 1, 2, \cdots, n$, where $\mu_{ij}$ is the number of links in the shortest path between $i$ and $j$, $\sum_{j=1}^{n} \mu_{ij}$ is the total number of links in the shortest paths between the country $i$ and all other nodes in the network. It is the reciprocal of the farness. In our problem, closeness centrality indicates how close a country is to all other countries in the Eurozone network. Thus, the more central a country $i$ is (high value of $c_i$), the closer it is to all other countries. If a country has strong closeness centrality, it is in a position, with its relationships, to exert a great influence on the others.

Betweenness and closeness centralities have been widely used in the literature of international trade network (Ercsey-Ravasz et al., 2012; Kuzubaş et al., 2014; Caballero, 2015; Ge et al., 2016; De Andrade and Rêgo, 2018; Petridis et al., 2020). If we consider the goods traded in the network, the distance between countries could have no clear interpretation because the goods traded between countries *a* and *b* might have nothing to do with the goods traded between countries *b* and *c* and, therefore, tell us nothing about the relation between *a* and *c*. But, if we look at the literature, this is not the proper way to interpret this concept. Countries in the top betweenness score play a critical role in the EMU network because any good will spread most efficiently through them into the rest of the network, and fast spread is also facilitated by the small value of the shortest path of them. Though a single, specific good may not necessarily follow the shortest paths in the EMU-network (e.g., it could be included into a more complex supply chain and sent on various routes), the small value of the shortest path length for node X is still an indicator of its proximity to almost all the nodes, guaranteeing fast spread on the network.

Betweenness considers how a given node plays as intermediary between any two nodes in the whole network. Countries that serve as middle points in a network structure, namely they receive and simultaneously send large flow of money, tend to have high value of betweenness centrality score and are called hubs. On the other hand, country with the largest score of betweenness has the largest load in the flow of international trade and plays a crucial role in the EMU network. Thus, the impact of countries with large betweenness centrality values to EMU trade is great.

Closeness centrality describes how "centered" a country X is in the network, according to the fact that the total distance between X and all the other countries is the shortest, i.e., it has the closest connections with other nodes. This means that the total distance between the country in the center of the network and all the other ones is the shortest, i.e., it is in a strong position in the trade process within EMU network. In this context, we introduce a normalized version of closeness centrality of a node *i*, named $\hat{c}_i$, in such a way the normalized score is bound between 0 and 1; it should be 0 if *i* is an isolate node, and 1 if it is directly connected to all others. Anyway, we are working with weighted graphs. The closeness value depends not only on which vertex is connected to, but also the edge weights. In fact, weighted closeness is not bound between 0 and 1 and, instead, can returns values greater than 1. We are looking for a way to normalize the closeness of a given node that is not the trivial one of dividing by the maximum score obtained on all nodes. Thus, denoted by $w_{tot}$ the sum of all the weights in the network and assuming:

1) to keep unchanged the relations that define the network;

2) to generate randomly the weights of each relation in such a way that their sum continues to be equal to $w_{tot}$;

3) to calculate the maximum weighted closeness among all the nodes, denoting it with

$$M = \max_{i \in |V|} c_i;$$

4) to repeat steps (2) and (3) K times and calculate $\bar{M} = \max_{k \in \{1,2,...,K\}} M_k$;

we wonder if the maximum weighted closeness (obtained for a given node i) from the original network is also maximum with respect to all those that can be generated by the K random networks or not. We then define the normalized version of (weighted) closeness centrality as follows[14]

$$\hat{c}_i = \frac{c_i}{\bar{M}} \quad for\ all\ i \in \{1,2,\dots,n\}$$

With the so-called link reciprocity (Garlaschelli and Loffredo, 2004) we focus on a peculiar type of correlation present in directed networks. It is the tendency of node pairs to form mutual connections between each other, i.e., the tendency of a node to form reciprocated social ties with others. In other words, with the reciprocity, we are interested in determining whether double links (i.e., double flow of money with opposite directions) occur between country pairs. Formally, the link reciprocity of a country/node $i$ is so defined:

$$lr_i = \sum_{j=1}^{n} w_{ij}\, w_{ji} \qquad (9)$$

for $i = 1, 2, \cdots, n$.

Another mechanism underlying network dynamics is clustering. Clustering describes the tendency of nodes, and therefore countries, to form ties within groups, where one's trading partners (from now on, *partners*) are connected to each other, too. This means that indirect connections between countries tend to become direct connections over time. In our problem, indirect connections can be two unconnected countries or weakly connected countries (i.e., countries with a lower flow of money between them), that trade with the same third party, or that are connected by a two-path. In a directed network, closure can be represented by transitive triplets and three-cycles (Block, 2015). A transitive triplet between three countries $i, j$ and $h$, is defined as a tie being present from $i$ to $j$, from $j$ to $h$, and a tie being present from $i$ to $h$, as depicted in Figure 4(a). It describes the tendency to form trading ties to partners of partners and mathematically is:

$$tt_i = \sum_{j,h} w_{ij} w_{ih} w_{jh} \qquad (9)$$

---

[14] See Appendix C for the algorithm that computes $\bar{M}$. Our results have been obtained by assuming $K = 10000$ runs.

for $i = 1, 2, \cdots, n$.

**Figure 4 – a) Transitive Triplet and b) Three-Cycle**

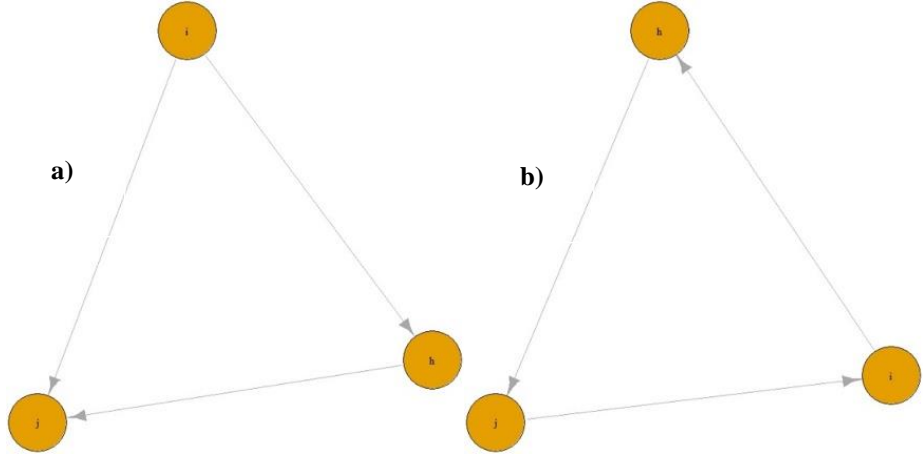

A three-cycle can be regarded as generalized reciprocity (in an exchange interpretation of the network). In a triad of countries $i$, $j$, and $h$, it describes the configuration where a tie from $i$ to $j$, a tie from $j$ to $h$, and a tie from $h$ to $i$ is present – see Figure 4(b). This configuration is like the transitive triplet, except that the direction of the tie from $i$ to $h$ is reversed and they coincide for undirected networks. It describes the tendency of country $i$ to form three-cycles. Formally, the three-cycle of a country/node $i$ is so defined:

$$c_i^{(3)} = \sum_{j,h} w_{ij} w_{hi} w_{jh} \tag{10}$$

for $i = 1, 2, \cdots, n$.

The out-degree popularity and in-degree popularity indicate preferential attachment processes (Snijders et al, 2010). If these measures are positive, countries with higher in-degree (for the in-degree popularity), or higher out-degree (for the out-degree popularity), are more attractive for others to start a trading tie to. They are defined as:

$$idp_i = \sum_j w_{ij} \sum_i w_{ij} \tag{11}$$

(for the in-degree popularity) and

$$odp_i = \sum_j w_{ij} \sum_j w_{ij} \tag{12}$$

(for the out-degree popularity).

*3.3 The EMU network*

As previously highlighted, German current account has been increasingly characterized by significant surpluses since the beginning of the EMU; this positive trend has represented an important break in relation to the previous negative dynamics which has distinguished all the Nineties (Figure 1). Paragraph 2 has introduced the main justifications behind this substantial change which has translated into persistent deficits in many countries of the European periphery. In any case, our main aim – among other things – is not to precisely identify the explanations of the German commercial rise but to understand if and to what extent Germany has represented a gravitational center of trade within the euro area. At this regard, we will take as a baseline scenario the year '1995', which can be considered as a proxy of the pre-euro trade situation. This scenario will be compared with the last available year of our series (2019), which allows us not to be worried about the health and the subsequent economic crisis happened in the beginning of 2020.

After having seen that the common currency has not been able to reduce trade asymmetry within the EMU, fostering on the contrary a polarization between (few) surplus countries and the others deficit countries, now we investigate which countries occupy the most influential positions in the trade network. To do that, we will exploit different centrality measures which have been explained in the previous paragraph and are reported in Table 2.[15] Generally speaking, within a certain year, each centrality measure is represented by an absolute value – the higher the value the higher the centrality. In order to show the results in a more convenient and meaningful way, each centrality measure has been expressed in two-fold ways. On the one side, it has been reported to the corresponding German value of the baseline scenario (1995); therefore, within a certain centrality measure, if a value is less than 1 (i.e., 0.5), this means that the relative country is characterized by a centrality which is half of the German benchmark; on the contrary, if a value is greater than 1 (i.e., 2), this translates into a more significant centrality which doubles that of the German baseline. On the other side, we estimated the overall amount of each centrality measure in absolute terms for each year and then we related the value of each country to the annual aggregate. In this way, it is possible to show the relative weight of the centrality of each country within a certain year.

Focusing on the first kind of normalization, it is worth noting that each index stresses the same outcome: Germany has always been the most important country in terms of centrality in trade flows and, over twenty-five years, it has been able to significantly increase its influence within the EMU network (Table 2a). This could appear a trivial result, but we have to stress at least two aspects. First of all, the increase in German centrality has been quite outstanding, regardless of the centrality measure taken as benchmark. Secondly, this outcome is even more significant if we compare it with Germany's import and export dynamics within the Eurozone: it is remarkable that German ability to increase its central role within the EMU network has gone hand in hand with a progressive decrease of at least ten percentage points in the share of imports and exports traded in the domestic market (Figure A.1 and A.2 in Appendix A). In this context, each centrality measure characterizes German trade predominance in a slightly different way. The first two centrality measures are related to the pervasiveness of a certain country in terms of exports (weighted out-degree centrality) and of imports (weighted in-degree centrality). This kind of index analyzes the network trade *rebus sic stantibus*, without considering any variation in the composition of the member states (this feature will be particularly important when we will consider the out- and in-popularity). From this point of view, Germany has more than doubled its influence on both side of trades. The reciprocity index introduces another interesting feature of the German predominance: since it describes the tendency of two nodes to form mutual connections, it is possible to note that only few countries show a not negligible influence. Among these, Germany has become a pole of attraction for mutual connections, quintupling its force of attraction – this interesting feature will be also confirmed by other kind of indicators. Transitive triplets and mostly three-cycles can be interpreted as extensions of the previous measure, which now take into consideration clusters of countries and their tendency to form ties within groups. In particular, it is confirmed that few member states are actually able to form this kind of connections, and Germany is – without any doubt – the most important actor. The last two centrality measures are related to preferential attachment processes. This implies that they measure the attractiveness of a certain country in relation to others in order to create a trading tie. As seen for the weighted out- and in-degree centralities, out-popularity refers to the export side while in-popularity to the import side. In this case, however, the network trade is not analyzed statically but wondering to what country a new member state would create a commercial link. At this regard, Germany confirms its supremacy and its tendency to become a unique attractor within the Eurozone.

---

[15] Table 2 only shows the extreme values of our period of analysis (1995 and 2019). In Appendix A, it is possible to find the centrality measures for the whole period (from Table A.1 to Table A.7).

**Table 2 – Centrality measures within the Eurozone**

**(a) Germany 1995 = 1**

|  | Out-degree | | In-degree | | Reciprocity | | Triplets | | Three-cycles | | Out-popularity | | In-popularity | |
|---|---|---|---|---|---|---|---|---|---|---|---|---|---|---|
|  | 1995 | 2019 | 1995 | 2019 | 1995 | 2019 | 1995 | 2019 | 1995 | 2019 | 1995 | 2019 | 1995 | 2019 |
| **Austria** | 0.1 | 0.4 | 0.2 | 0.5 | 0.1 | 0.4 | 0.0 | 0.5 | 0.1 | 0.6 | 0.0 | 0.1 | 0.3 | 1.3 |
| **Belgium** | 0.4 | 1.0 | 0.4 | 0.9 | 0.3 | 1.2 | 0.5 | 4.7 | 0.4 | 4.0 | 0.2 | 1.0 | 0.7 | 3.2 |
| **Cyprus** | - | - | - | - | - | - | - | - | - | - | - | - | - | - |
| **Estonia** | - | - | - | - | - | - | - | - | - | - | - | - | - | - |
| **Finland** | 0.1 | 0.1 | - | 0.1 | - | - | - | 0.1 | - | 0.1 | - | - | 0.1 | 0.4 |
| **France** | 0.6 | 1.1 | 0.7 | 1.5 | 0.6 | 2.0 | 0.7 | 3.8 | 0.8 | 5.7 | 0.4 | 1.1 | 0.8 | 3.0 |
| **Germany** | **1.0** | 2.2 | **1.0** | 2.1 | **1.0** | 4.1 | **1.0** | 9.0 | **1.0** | 8.3 | **1.0** | 5.0 | **1.0** | 4.6 |
| Greece | - | 0.1 | 0.1 | 0.1 | - | - | - | - | - | - | - | - | - | 0.1 |
| Ireland | 0.1 | 0.3 | - | 0.1 | - | - | - | 0.4 | - | 0.2 | - | 0.1 | 0.1 | 0.9 |
| Italy | 0.5 | 0.9 | 0.5 | 1.0 | 0.3 | 1.1 | 0.5 | 3.3 | 0.5 | 3.6 | 0.2 | 0.9 | 0.8 | 2.8 |
| Latvia | - | - | - | - | - | - | - | - | - | - | - | - | - | - |
| Lithuania | - | 0.1 | - | 0.1 | - | - | - | - | - | - | - | - | - | 0.1 |
| Luxembourg | - | - | - | 0.1 | - | - | - | - | - | - | - | - | - | 0.2 |
| Malta | - | - | - | - | - | - | - | - | - | - | - | - | - | - |
| Netherlands | 0.5 | 1.2 | 0.4 | 0.9 | 0.3 | 1.8 | 0.5 | 5.8 | 0.4 | 4.2 | 0.2 | 1.4 | 0.7 | 3.9 |
| Portugal | 0.1 | 0.2 | 0.1 | 0.3 | - | 0.1 | - | 0.1 | - | 0.2 | - | - | 0.1 | 0.5 |
| Slovakia | - | 0.2 | - | 0.1 | - | - | - | 0.1 | - | 0.1 | - | - | - | 0.6 |
| Slovenia | - | 0.1 | - | 0.1 | - | - | - | - | - | - | - | - | - | 0.3 |
| Spain | 0.2 | 0.7 | 0.3 | 0.8 | 0.1 | 0.6 | 0.1 | 1.8 | 0.2 | 2.0 | 0.1 | 0.5 | 0.3 | 2.0 |

**(b) % on Total (intra-year)**

|  | Out-degree | | In-degree | | Reciprocity | | Triplets | | Three-cycles | | Out-popularity | | In-popularity | |
|---|---|---|---|---|---|---|---|---|---|---|---|---|---|---|
|  | 1995 | 2019 | 1995 | 2019 | 1995 | 2019 | 1995 | 2019 | 1995 | 2019 | 1995 | 2019 | 1995 | 2019 |
| Austria | 3.9 | 4.3 | 5.0 | 5.3 | 2.8 | 3.9 | 1.2 | 1.7 | 1.6 | 2.1 | 1.0 | 1.3 | 5.1 | 5.6 |
| Belgium | 11.9 | 11.7 | 10.3 | 10.6 | 9.7 | 10.7 | 14.2 | 15.8 | 12.6 | 13.9 | 9.2 | 9.9 | 13.3 | 13.3 |
| Cyprus | - | 0.1 | 0.2 | 0.3 | - | - | - | - | - | - | - | - | - | - |
| Estonia | 0.1 | 0.4 | 0.2 | 0.5 | - | - | - | - | - | - | - | - | 0.1 | 0.2 |
| Finland | 1.7 | 1.4 | 1.3 | 1.4 | 0.2 | 0.2 | 0.2 | 0.2 | 0.2 | 0.2 | 0.2 | 0.2 | 1.8 | 1.6 |
| France | 16.8 | 12.4 | 19.0 | 17.4 | 22.8 | 17.6 | 20.0 | 12.8 | 23.6 | 19.6 | 18.4 | 11.2 | 16.5 | 12.4 |
| Germany | 26.7 | 25.9 | 25.9 | 23.6 | 36.5 | 35.4 | 29.7 | 30.3 | 29.4 | 28.6 | 46.3 | 48.4 | 19.7 | 19.4 |
| Greece | 0.7 | 0.7 | 1.7 | 1.3 | 0.1 | 0.1 | - | - | 0.1 | 0.1 | 0.0 | 0.0 | 0.8 | 0.6 |
| Ireland | 2.1 | 3.3 | 0.9 | 1.3 | 0.1 | 0.3 | 0.4 | 1.3 | 0.2 | 0.6 | 0.3 | 0.8 | 2.3 | 3.6 |
| Italy | 13.3 | 10.8 | 12.3 | 11.1 | 12.6 | 9.5 | 15.2 | 11.2 | 14.9 | 12.2 | 11.5 | 8.4 | 14.8 | 11.7 |
| Latvia | 0.1 | 0.3 | 0.1 | 0.5 | - | - | - | - | - | - | - | - | 0.1 | 0.2 |
| Lithuania | 0.1 | 0.6 | 0.2 | 0.7 | - | - | - | - | - | - | - | - | 0.1 | 0.4 |
| Luxembourg | 0.6 | 0.5 | 1.1 | 0.9 | 0.1 | - | - | - | 0.1 | 0.1 | - | - | 0.7 | 0.6 |
| Malta | 0.2 | 0.1 | 0.2 | 0.3 | - | - | - | - | - | - | - | - | 0.2 | 0.1 |
| Netherlands | 12.5 | 13.9 | 10.3 | 10.6 | 11.3 | 15.6 | 14.7 | 19.5 | 12.1 | 14.6 | 10.1 | 14.0 | 14.4 | 16.2 |
| Portugal | 1.8 | 2.1 | 2.6 | 3.0 | 0.3 | 0.7 | 0.3 | 0.4 | 0.4 | 0.6 | 0.2 | 0.3 | 1.9 | 2.1 |
| Slovakia | 0.4 | 2.1 | 0.5 | 1.6 | - | 0.4 | - | 0.5 | - | 0.4 | - | 0.3 | 0.5 | 2.6 |
| Slovenia | 0.6 | 1.0 | 0.8 | 1.0 | - | 0.1 | - | 0.1 | - | 0.1 | - | 0.1 | 0.8 | 1.1 |
| Spain | 6.5 | 8.3 | 7.6 | 8.5 | 3.5 | 5.5 | 3.8 | 6.1 | 4.7 | 7.0 | 2.7 | 5.0 | 6.8 | 8.4 |

*Note:* the hyphens represent a situation in which a country shows a negligible relative centrality. Definitions of the different centrality measures can be found in Paragraph 3.2. Appendix A reports the centrality measures for the whole period (from Table A.1 to Table A.7)
*Source:* elaborations on *Observatory of Economic Complexity* (*OEC*) data

Once we normalized each centrality value in relation to the annual aggregate, we estimate how the relative weight of each country – in terms of the corresponding centrality – has changed over time (Table 2b). In this case, Germany's growing centrality may appear more moderate, but it must be contextualized from a double point of view. First, Germany has been always able to maintain its share of centrality. This result appears even more significant if we link it to the second aspect: a major part of the countries has increased its influence on the overall centrality, and this has mainly happened to the detriment of France and Italy. Moreover, as previously seen, Germany's centrality in the EMU network has been kept within a trade flow that has largely developed towards countries outside the monetary union (Figures A.1 and A.2 in Appendix A).

Another way to understand how the single currency has influenced the trade flow within the Eurozone is to look at a second kind of characterization related to the commercial path among member states. As previously highlighted, although it may appear counterintuitive due to the lack of input-output data, this way of representing the network has been largely and meaningfully used in this kind of literature. In particular, we will take into consideration three different aspects: first, how close a country is to all other countries in the Eurozone network (*closeness*); secondly, the ability of a country to be an intermediary between any two nodes, which helps identify what countries play a bridging role within a network (*betweenness*); finally, how many times a country directly sends/receives the goods to/from another country without going through other states (*direct exporter*/*direct importer*) (Table 3). As opposed to other centrality measures, closeness reveals how the EMU network is characterized by a lot of countries which plays an important role in the trade flows. In any case, Germany has always been the most central countries over the whole period, increasing its relative position of influence. The second aspect underlines the fundamental bridging role of Germany within Eurozone network: even if the intra-year trend denotes a slight decrease (69.9 in 1995 and 62.3 in 2019), Germany continues to play a crucial role in the EMU. In this context, this slight reduction of its influence as an intermediary can be explained by looking at the times it has been a direct exporter and importer: Germany is one of the few Eurozone countries which has increased its overall role as a direct exporter and importer.

**Table 3 – Shortest paths within the Eurozone**

|  | Closeness | | Betweenness | | | | Direct exporter | | Direct importer | |
| --- | --- | --- | --- | --- | --- | --- | --- | --- | --- | --- |
|  | *Normalised* | | *Germany 1995 = 1* | | *% on total (intra-year)* | | | | | |
|  | 1995 | 2019 | 1995 | 2019 | 1995 | 2019 | 1995 | 2019 | 1995 | 2019 |
| **Austria** | 0.58 | 0.80 | - | - | 0.0 | 0.1 | 1 | 1 | 1 | 1 |
| **Belgium** | 0.59 | 0.82 | - | 0.1 | 0.4 | 2.1 | 4 | 4 | 3 | 4 |
| **Cyprus** | 0.12 | 0.17 | - | - | 0.0 | 0.0 | 1 | 1 | 1 | 1 |
| **Estonia** | 0.19 | 0.31 | - | - | 0.0 | 0.0 | 1 | 2 | 1 | 4 |
| **Finland** | 0.51 | 0.64 | - | - | 0.3 | 0.2 | 2 | 2 | 2 | 2 |
| **France** | 0.60 | 0.83 | 0.2 | 0.6 | 16.3 | 19.3 | 4 | 4 | 4 | 4 |
| **Germany** | 0.61 | 0.87 | **1.0** | 1.8 | 69.9 | 62.3 | 13 | 14 | 15 | 13 |
| **Greece** | 0.42 | 0.47 | - | - | 0.2 | 0.2 | 1 | 2 | 2 | 2 |
| **Ireland** | 0.48 | 0.70 | - | - | 0.0 | 0.0 | 1 | 2 | 1 | 1 |
| **Italy** | 0.60 | 0.82 | 0.2 | 0.2 | 11.7 | 8.0 | 4 | 3 | 3 | 4 |
| **Latvia** | 0.12 | 0.26 | - | - | 0.0 | 0.0 | 1 | 2 | 1 | 3 |
| **Lithuania** | 0.17 | 0.39 | - | - | 0.0 | 0.1 | 1 | 3 | 1 | 2 |
| **Luxembourg** | 0.39 | 0.45 | - | - | 0.0 | 0.0 | 1 | 1 | 1 | 1 |
| **Malta** | 0.25 | 0.19 | - | - | 0.0 | 0.0 | 1 | 1 | 1 | 1 |
| **Netherlands** | 0.59 | 0.84 | - | 0.2 | 1.1 | 5.3 | 2 | 2 | 2 | 2 |
| **Portugal** | 0.49 | 0.65 | - | - | 0.0 | 0.0 | 1 | 1 | 1 | 1 |
| **Slovakia** | 0.37 | 0.71 | - | - | 0.0 | 0.0 | 1 | 1 | 1 | 1 |
| **Slovenia** | 0.41 | 0.55 | - | - | 0.0 | 0.0 | 1 | 1 | 1 | 1 |
| **Spain** | 0.57 | 0.79 | - | 0.1 | 0.2 | 2.4 | 3 | 4 | 2 | 3 |

*Note:* the column 'direct exporter' identifies the number of times each country directly sends the goods to another country without going through other states; the column 'direct importer' exactly reports the same information but in relation to imports. The hyphens represent a situation in which a country shows a negligible relative centrality. Definitions of the different centrality measures can be found in Paragraph 3.2. Appendix A reports the centrality measures for the whole period (Tables A.8 and A.9)
*Source:* elaborations on *Observatory of Economic Complexity* (*OEC*) data

## 4. Conclusions and policy implications

Trade imbalances within a monetary union can be considered an unfair outcome: in the absence of a single central bank, states running a trade deficit could implement autonomous monetary policy decisions – i.e., if a deficit country devalues its currency, it makes its exports more attractive and its imports less attractive – to improve their trade balance; on the contrary, if two or more countries share the same currency, they are not able to use monetary policy to reduce trade imbalances with other members (Meyer, 2021). In other words, being part of a monetary union such as the EMU implies the loss of any control over the exchange rate, which could normally help balance national trade flows. Consequently, the rebalancing process among member states can be more difficult than outside a monetary union. This has led in the past to a wide debate within the European institutions on the absence of proper rules which could prevent such imbalances and the adoption of the Macroeconomic Imbalances Procedure (MIP) within the EMU countries at the end of 2011 precisely meets this need. In any case, it is important to stress the attention on the fact that, from a theoretical perspective, trade imbalances are potentially dangerous both for deficit and surplus countries. On the one side, as previously highlighted, deficit countries must finance their imports by developing new debt. From this point of view, trade balance can also be seen as a better index of macroeconomic sustainability of a Eurozone member rather than its budget balance or the level of public debt (Bénassy-Quéré, 2017). On the other side, trade imbalances may represent an issue also for surplus countries, whose export-led growth model could collapse in front of multiple defaults of deficit countries.

In this context, fiscal policy and the unconditional presence of a lender of last resort can stabilize output within a monetary union but they cannot eliminate the causes of trade imbalances. This implies that existing discrepancies in export competitiveness should directly be tackled, and this should happen both at national and at a European level (Gräbner et al., 2021). In recent years, in fact, given that a nominal unilateral devaluation is impossible within a single currency area, the attempting to lower external deficits has been primarily achieved through internal devaluations (Arachi and Assisi, 2021), and this policy has showed large economic and social costs (Stockhammer and Sotiropoulos, 2014). If wage moderation has theoretically been often seen as a way-out economic crisis for countries like Greece, from an empirical point of view this does not seem to have been the case: wage flexibility has not been capable of solving divergences in the levels of competitiveness and of current account positions across Europe (Stockhammer, 2011).

By applying network analysis to trade flows in goods within the Eurozone and covering a period of twenty-five years (1995-2019), we have used a series of centrality measures in order to identify the most influential countries within the single currency area. All the related outcomes are incontrovertible, highlighting the fundamental role of Germany as a European hub within the Eurozone trade network over time. In particular, German predominance has noticeably grown from the birth of the euro onwards, emphasizing a polarization phenomenon in relation to trade imbalances. Twenty years of the single currency seem to have deeply changed trade flows among member states of the Eurozone and Germany has intercepted these changes, becoming one of the most important export-led country. We have also seen that the single market encouraged – rather than reduced – trade imbalances: the 'Eurozone current account core-periphery dualism' has been fostered by the adoption of the single currency and this has been associated by a relevant increase in trade asymmetry.

These results appear unsustainable, creating a '*beggar-thy-neighbour*' situation where one country tries to improve its economic condition by worsening the economic issues of other countries. Expansionary policies on the demand side – i.e., higher wage growth but also an expansionary fiscal policy – seem more necessary than ever in those countries constantly in trade surplus – and with balanced public budgets – such as Germany.

**Data Availability Statement**

The data that support the findings of this study are available from the corresponding author upon reasonable request.